\documentclass{aa}
\usepackage{natbib}  

\usepackage{graphicx}
\usepackage{color}
\usepackage{txfonts}
\usepackage[colorlinks=true,linkcolor=blue,citecolor=blue]{hyperref}
\usepackage{lscape}
\usepackage{amsmath}
\usepackage{amsfonts}
\usepackage{amssymb}
\begin{document}

   \title{Timing the warm absorber in  NGC 4051}

   \author{C. V. Silva
          \inst{1,2}\fnmsep\thanks{c.v.dejesussilva@uva.nl}
          \and
           P. Uttley\inst{1}
          \and
          E. Costantini\inst{2}
          }

   \institute{Anton Pannekoek Institute for Astronomy, University of Amsterdam,
              Science Park 904, 1098 XH Amsterdam, The Netherlands\\
         \and
             SRON, Netherlands Institute for Space Research, Sorbonnelaan 2, 3584 CA Utrecht, The Netherlands\\
             }

   \date{}

 \abstract
{We investigated, using spectral-timing analysis, the characterization of highly ionized outflows in Seyfert galaxies, the so-called warm absorbers. Here, we present our results on the extensive $\sim$600 ks of XMM-Newton archival observations of the bright and highly variable Seyfert 1 galaxy NGC 4051, whose spectrum has revealed a complex multicomponent wind. Making use of both RGS and EPIC-pn data, we performed a detailed analysis through a time-dependent photoionization code in combination with spectral and Fourier spectral-timing techniques. The source light curves and the warm absorber parameters obtained from the data were used to simulate the response of the gas due to variations in the ionizing flux of the central source. The resulting time variable spectra were employed to predict the effects of the warm absorber on the time lags and coherence of the energy dependent light curves. We have found that, in the absence of any other lag mechanisms, a warm absorber with the characteristics of the one observed in NGC 4051, is able to produce soft lags, up to 100 s, on timescales of $\sim \text{hours}$. The time delay is associated with the response of the gas to changes in the ionizing source, either by photoionization or radiative recombination, which is dependent on its density. The range of radial distances that, under our assumptions, yield longer time delays are distances $r\sim0.3-1.0 \times 10^{16}$ cm, and hence gas densities $n\sim0.4-3.0\times10^{7}\ \text{cm}^{-3}$. Since these ranges are comparable to the existing estimates of the location of the warm absorber in NGC 4051, we suggest that it is likely that the observed X-ray time lags may carry a signature of the warm absorber response time, to changes in the ionizing continuum. Our results show that the warm absorber in NGC 4051 does not introduce lags on the short time-scales associated with reverberation, but will likely modify the hard continuum lags seen on longer time-scales, which in this source have been measured to be on the order of $\sim 50$ s. Hence, these results highlight the importance of understanding the contribution of the warm absorber to the AGN X-ray time lags, since it is also vital information for interpreting the lags associated with propagation and reverberation effects in the inner emitting regions.}

   \maketitle
%
%
\section{Introduction}
Active galactic nuclei (AGN) are powered by accretion onto a supermassive black hole ($10^{6}-10^{9}\ \text{M}_\odot$). Outflowing events are often also associated with AGN. This ejection of matter and energy, if powerful enough, may affect the the surrounding environment of the AGN and even disturb the evolution of the host galaxy or the cluster hosting the AGN, a phenomenon often termed as AGN feedback \citep[e.g.][]{dimatteo2005,hardcastle2007,fabian2012,crenshaw2012}. The impact of the outflows on the surrounding environment is a function of their distance to the central source, the column density of the outflowing gas, and is highly dependent on the outflowing velocities. Hence, it is crucial to investigate the physical properties of the gas in order to assess its importance for feedback. However, while the column density and the outflowing velocity of the gas are inferred from observations, it is not possible to directly estimate the distance of the gas to the central source. \par
In the case of Seyfert 1 galaxies, $60\%$ show the presence of highly ionized outflowing absorbing gas, rich in metals \citep{crenshaw1999}. The outflowing material, usually called a warm absorber, is remarkably complex in its structure, spanning a wide range in ionization parameter, outflowing velocities, and column densities. These outflows have been detected both in the UV and in the X-ray spectra of Seyfert 1 galaxies, through a composite set of absorption lines \citep[for a review see][]{crenshaw2003}. Determining the radial location of the outflows yields valuable information for the study of AGN feedback. Yet characterizing the spatial location of the warm absorbers is not trivial. A common approach to determine the distance of the absorber to the central source is by measuring the density $n$ of the gas using sensitive absorption lines, and deriving the distance $r$ through the ionization parameter $\xi$, where $\xi=L_\text{ion}/nr^{2}$, \citep[e.g.][]{kraemer2006,arav2008}. This method is usually successful for UV data, where these lines are more commonly found, but it is not very effective to study warm absorbers in the X-rays \cite[e.g.][]{kaastra2004}. There we lack the instrumental sensitivity necessary to achieve such measurements. Alternatively, monitoring the response of the gas to changes in the ionizing continuum leads to an estimation of a recombination timescale, which is a function of the electron density. This approach has been applied through time resolved spectroscopy studies and time-dependent photoionization models \citep{behar2003,reeves2004,krongold2007,steenbrugge2009,kaastra2012}. Time resolved spectroscopy in the X-rays suffers from the problem that the involved timescales may be on the order of minutes to hours, which yields limited photon counts, per time and energy bin. The low signal to noise issue can be avoided by studying the statistical properties of variability instead, through the use of Fourier spectral-timing techniques. \par
Fourier spectral-timing techniques have been applied to the study of AGN X-ray light curves for more than a decade. It has been found in many sources that soft and hard X-ray photons behave differently on different timescales. For long timescales, the hard photons arrive with a time delay compared to the soft photons. On the contrary, on short timescales the soft photons lag behind the hard photons. The complex time-scale-dependent lags are associated with different physical processes and disentangling them is essential to understand the innermost emitting regions in AGN. The soft lag associated with short timescales has been explained through reverberation from reflected emission \citep{fabian2009}. This scenario is supported by the recently found Fe K$\alpha$ lags \citep{zoghbi2012,kara2013}. Furthermore, it has been suggested that fluctuations in the accretion flow propagate inwards, so that the outermost soft X-rays respond faster than the innermost hard X-rays, which could explain the hard lag seen on long timescales \citep{kotov2001,arevalo2006}. Most of the AGN with such timing properties, are Seyfert 1 galaxies. Since most Seyfert 1 galaxies show the presence of a warm absorber, our goal in this paper is to explore the possible contribution of the warm absorber to the observed X-ray time lags, due to the delayed response of the gas to the continuum variations.  \par
In this work we expand on the method of \cite{kaastra2012}, using a time-dependent photoionization model to examine the response of the gas to changes in the ionizing continuum. NGC 4051, a relatively nearby Seyfert 1 galaxy ($z\approx0.002$), is an ideal candidate for this study. This source is not only bright (with a luminosity in the 2-10 keV band of $L_\text{X}\sim2.1\times10^{41}\ \text{erg}\ \text{s}^{-1}$ and a corresponding observed flux of $f_\text{X}\sim1.6\times10^{-11}\ \text{erg}\ \text{s}^{-1}\ \text{cm}^{-2}$, in the data used here), but also highly variable \citep{mchardy2004} and has been associated with a complex multicomponent warm absorber \citep{steenbrugge2009,pounds2011_1}. NGC 4051 also shows the characteristic X-ray time lags found in other AGN \citep{alston2013}, with a hard lag at low frequencies and a soft lag at high frequencies. Furthermore, NGC 4051 has an extensive series of XMM-Newton archival observations. In section \ref{section2}, we present the methods used in the reduction and processing of the raw data products, as well as the procedures used to extract the light curves and spectra. We simulate the response of the complex warm absorber observed in the energy spectra of NGC 4051, to changes in the luminosity of the central source, and its dependence on radial distance and density, in section \ref{section3}. We further investigate in section \ref{section4} whether a non-equilibrium gas phase, which results in a delayed response of the gas to the variations on the central source, could produce a time delay of the most absorbed X-ray bands relative to the broad X-ray ionizing continuum. We finally compare our simulations to the real data in section \ref{section5} and present our conclusions in section \ref{section6}.\par
Throughout this work we use a flat cosmological model with $\Omega_{m}=0.3$, $\Omega_{\Lambda}=0.7$ and $\Omega_{r}=0.0$, together with a cosmological constant $\text{H}_{0}=70\ \text{km}\ \text{s}^{-1}\ \text{Mpc}^{-1}$. For the spectral modelling in this paper we have assumed a Galactic column density of $N_\text{H}=1.15\times10^{20}\ \text{cm}^{-2}$ \citep{kalberla2005}. The errors quoted in this paper are $1\sigma$ errors, unless otherwise stated.
%
\section{Observations and data reduction}
 \label{section2}
NGC 4051 was extensively observed by XMM Newton. We make use of the 15 observations from 2009, a total of $\sim570$ ks of data \citep[see][for details]{vaughan11}. \par
 \subsection{RGS and EPIC-pn light curves}
 \begin{figure*}
 \centering
  \includegraphics[width=\textwidth]{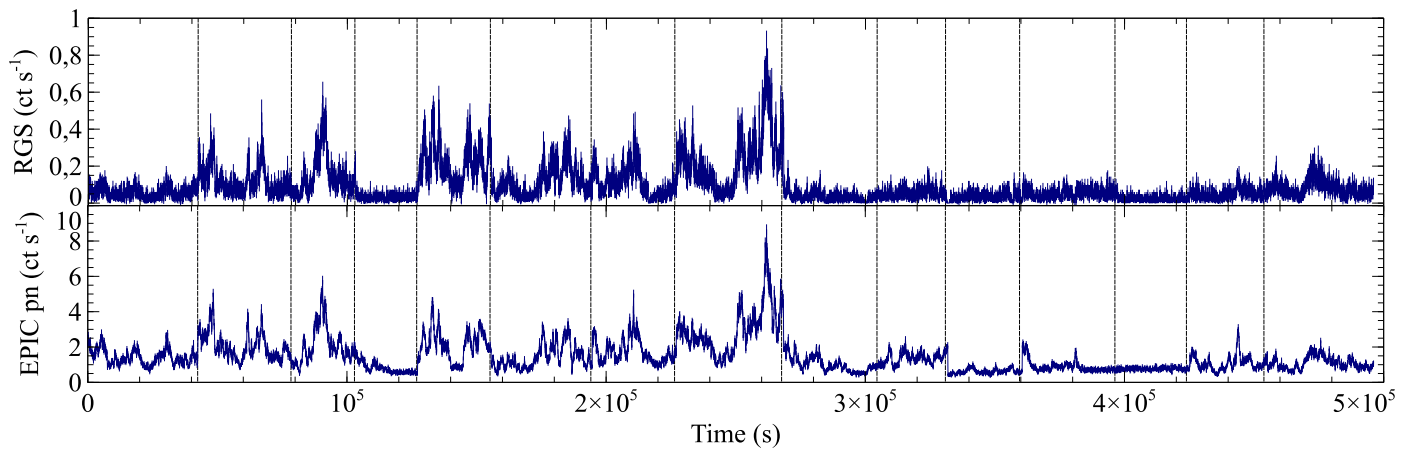}
  \caption{RGS light curves for the 15-17$\AA$ band (top) and EPIC-pn light curves for 1.5-10 keV (bottom). The dashed lines separate each individual observation.}
  \label{lc}
\end{figure*}
The Observation Data Files (ODFs) were reprocessed using the \textit{XMM-Newton} Science Analysis System (\textsc{sas} v.13.5) to generate calibrated event lists for RGS and for EPIC-pn, in which the conditions \texttt{PATTERN<=4} and \texttt{FLAG=}0 were used. For the purposes of this work, RGS and EPIC-pn event lists were filtered through a common Good Time Intervals (GTI) file, created with the task \textsc{mgtime} from HEASOFT (v. 6.15), to only include time intervals in which RGS and EPIC-pn were simultaneously observing. For EPIC-pn we extracted source and background event lists by selecting circular regions with a radius of 20 arcsec. All the observations were inspected for high particle background flaring and this was properly accounted in the filtering process.\par
RGS background subtracted light curves were extracted with the SAS task \textsc{rgslccorr}, for the two RGS instruments combined and first order selected, with a time binsize of 100 s. EPIC-pn background subtracted light curves were obtained by using the RGS light curves as template for start and stop times, and for binsize, such that both RGS and EPIC-pn light curves are evenly sampled in time. This was done for each observation separately (see Fig.~\ref{lc}). The light curves were also corrected for short exposure losses by linear interpolation of the nearest good data points in each side, adding appropriate Poisson noise \citep[as seen in][]{vaughan11}.
\subsection{RGS and EPIC-pn spectra}
RGS spectra (source and background) together with Response Matrix Files (RMFs) were extracted for each observation by using SAS, selecting first order. The averaged RGS spectrum was obtained using the SAS task \textsc{rgscombine} which combines the spectra of all observations and creates a single RMF. Source and background EPIC-pn spectra were extracted with SAS for each observation, simultaneously with the corresponding RMFs and Ancillary Response Files (ARFs). Using HEASOFT (v.6.15), we produced a time-averaged EPIC-pn spectrum both for the source and for the background, adding the individual spectra with the ftool   \textsc{mathpha}. To combine the response matrix and effective area files we made use of the ftools \textsc{addrmf} and \textsc{addarf} respectively. 
%
%
\section{Time-dependent photoionization model}
 \label{section3}
The main goal in this paper is to study the possible effects of a warm absorber on the spectral-timing data from NGC 4051. In order to do so, we perform a detailed study of the response of the outflowing gas to the fast changes in the ionizing continuum. Using a time-dependent photoionization model we investigate the variations in the spectrum, due to a complex warm absorber illuminated by a variable continuum source.  The correct modelling of the spectrum, with time, will allow us to generate light curves, which depend on the flux history of the source.\par 
For photoionization, a variable X-ray source will have an effect on the ionization balance of the gas. The outflowing gas, which is rich in metals, responds to the changes in the ionizing flux in the following manner. When the flux of the ionizing source increases, the gas becomes more ionized. In the same way, when the flux of the ionizing source decreases, the gas recombines. In order to monitor the time evolution of the ion concentrations in the gas, it is necessary to solve the time-dependent ionization balance equations. As such, the relative density $n_{\text{X}^{i}}$ of ion $i$, of a certain species $\text{X}$, varies with time as a function of the electron density of the gas, $n_\text{e}$, the recombination rate from stage $i+1$ to $i$ (given by the recombination coefficient, $\alpha_{\text{rec},\text{ X}^{i}}$, times the electron density), and the ionization rate from stage $i$ to $i+1$, $I_{\text{X}^{i}}$. This time dependence is given by \cite{krolik1995}, and can be written as
 \begin{equation} \label{eqn:time_con}
\frac{\text{d}n_{\text{X}^{i}}}{\text{d}t}=-n_{\text{X}^{i}}n_\text{e}\alpha_{\text{rec},\text{X}^{i-1}}-n_{\text{X}^{i}}I_{\text{X}^{i}}+n_{\text{X}^{i+1}}n_\text{e}\alpha_{\text{rec},\text{X}^{i}}+n_{\text{X}^{i-1}}I_{\text{X}^{i-1}}
\end{equation}
in which Auger ionization, collisional ionization, and three-body recombination were neglected. Equation (\ref{eqn:time_con}) is thus defined by the sum of the destruction rate and formation rate of each ion, only due to photoionization and radiative recombination. \par
If this response is instantaneous, the gas is in photoionization equilibrium with the ionizing continuum at all times. However, this has been shown to not always be the case and so the processes of ionization and recombination of the gas can be associated with ionization and recombination timescales. Ionization and recombination timescales correspond to the the time it takes for the gas to reach equilibrium with the ionizing continuum, for increasing or decreasing flux phases respectively. Following \cite{nicastro1999}, this timescale can be approximated as
\begin{equation}
t_{\text{eq}}^{\text{X}^{i},\text{X}^{i+1}}\sim\left(\frac{1}{\alpha_{\text{rec},\text{X}^{i}}n_\text{e}}\right)\times\left[\frac{1}{(\alpha_{\text{rec},\text{X}^{i-1}}/\alpha_{\text{rec},\text{X}^{i}})+(n_{\text{X}^{i+1}}/n_{\text{X}^{i}})}\right]
\end{equation}
In this situation, the gas will reach equilibrium with the ionizing continuum after a time delay, $t_\text{eq}$. Since the equilibrium timescale shows a dependence on the electron density, measuring this time delay would allow us to constrain the density of the gas. The same time delay could also produce an overall delay effect of the more absorbed energy bands relative to the continuum, which could affect the X-ray lag measurements in these systems. We investigate this possibility in detail in section \ref{section4}.\par 
Another possibility for the behaviour of this gas is that its response time is very long compared to the typical variability timescales of the source. In this case, the gas is in a steady state, but never reaches equilibrium with the ionizing continuum \citep{krolik1995}. At best, the densities of the ions remain fairly constant around their mean values, with the ionization and recombination rates corresponding to the values associated with the mean flux of the source over time. \par
All this complex behaviour can be assessed by solving equation (\ref{eqn:time_con}) and studying the time dependence of the ions. Our approach to obtain the time-dependent ion concentrations follows that of \cite{kaastra2012}. The whole method is described bellow, including some results from the simulations we performed with the calculated concentrations.
\subsection{The ionizing continuum}
\label{subsection31}
NGC 4051 is highly variable, showing large variations in flux over relatively short timescales, on the order of $\sim$ few hours \citep{mchardy2004,vaughan11}. This intrinsic rapid variability, together with its complex absorbed spectrum, makes NGC 4051 an ideal candidate for this study. 
The ionizing continuum used to obtain the ionization balance takes as reference the time-averaged spectral energy distribution (SED) of the 15 observations of NGC 4051 mentioned in section \ref{section2}. It is important to note here that for the purpose of our study we assume the ionizing continuum only changes in X-ray flux normalisation. The shape of the SED that we consider here is kept constant with time. The reason is to not include any other source of lags due to the continuum variations. In this way, the light curves we generate are only affected by the variable warm absorber, excluding any effects due to changes in the spectral shape with time, which naturally introduces time delays between different energy bands. For details see section \ref{subsection35}. \par
The X-ray shape of the SED is taken from the time-averaged spectrum of EPIC-pn. We have performed a phenomenological fit using the spectral fitting package SPEX v.2.05 \citep{kaastra1996}, to obtain the time-average shape of the broad X-ray spectrum ($0.5-10$ keV). The model to describe the overall shape of the continuum consists of a steep power law ($\Gamma\sim3.02$) with a spectral break at $\sim1.47$ keV, which leads to a harder spectrum ($\Gamma\sim1.75$). To better model the spectrum in the soft band, where strong absorption features are detected in the RGS data, we add a set of warm absorption models. These absorption models result from fitting the time-averaged RGS spectrum using the \textsc{xabs} model from SPEX, corresponding to absorption by a slab of material in photoionization equilibrium. In this step we use the default ionization balance in SPEX, set by running CLOUDY v.13.01 \citep{ferland2013} with a standard SED shape from NGC 5548. This is a good approximation since at this stage we only aim to constrain the overall shape of the EPIC-pn spectrum, and the changes that the appropriate ionization balance  introduces are small at the available resolution. Furthermore, a positive Gaussian profile centered at $\sim6.44$ keV, with width $\sigma\sim160$ eV, was added to account for the Fe K feature visible in the spectrum. \par
\begin{figure}[top]
  \centering
   \includegraphics[width=\hsize]{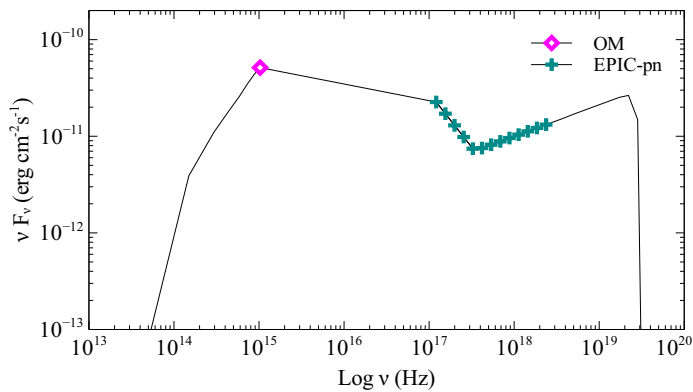}
     \caption{Spectral energy distribution of NGC 4051 for the averaged EPIC-pn and OM data.}
      \label{fig:sed}
 \end{figure}

To complete the time-averaged SED, we made use of the OM data, also available for these observations. The OM data were collected in Imaging mode in the UVW1 filter, corresponding to a central wavelength of $2910 \ \AA$ \citep[see][for more details]{alston2013_2}. The OM count rates were extracted with SAS task  \textsc{omchain} and converted to fluxes by using the recommended conversion factor. We have taken the average flux of the 15 observations and use this value as the time-averaged flux at $2910 \ \AA$. Beyond the OM data, we make use of the the default AGN continuum in CLOUDY, which is characterized by relatively low luminosity at longer wavelengths \citep{mathews1987}. The SED is completed by expanding the continuum above 10 keV, adding an artificial cut-off at $\sim$150 keV. The broad-band SED is presented in Fig. \ref{fig:sed}.\par
\subsection{The multicomponent outflowing gas}
\label{subsection32}
To accurately model the time evolution of the ion concentrations in the gas, it is first necessary to model the RGS spectrum, accounting for the numerous absorption features, observable in the $10-36 \ \AA$ range. Considering that the continuum spectral shape is represented here in such a narrow energy band, we opted to choose a phenomenological continuum model, that can better describe only the soft band. As such, we fit the underlying continuum of the time-averaged RGS spectrum with a blackbody emission component ($T\sim0.15$ keV) in addition to a power-law model. The power-law has an index $\Gamma\sim3.1$. To account for the complex absorption features observed, we use the SPEX photoionization absorption model  \textsc{xabs}. The ionization balance is calculated with CLOUDY, using the previously constructed SED (see Fig. \ref{fig:sed}). Our best fit results in the detection of four distinct photoionized components, summarized in table \ref{table:wa_comp}.\par
\begin{figure}[t]
   \centering
   \includegraphics[width=\hsize]{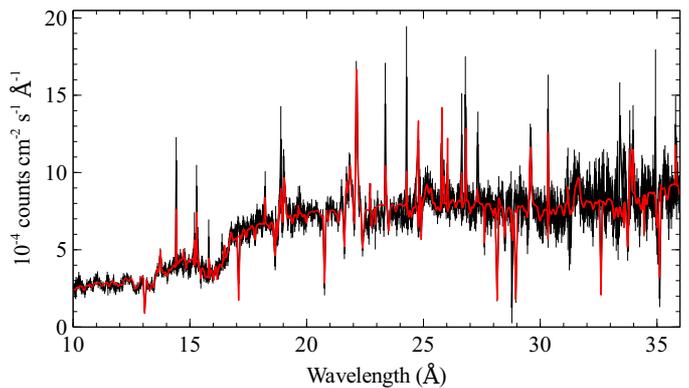}
     \caption{Best fit model (in red) of the time averaged RGS spectrum (in black) of NGC 4051.}
       \label{fig:rgs_spectrum}
  \end{figure}

 \begin{table}[b]
\caption{Best fit parameters of the RGS spectrum for the multicomponent photoionized outflow.}             
\label{table:wa_comp}      
\centering                          
\begin{tabular}{c c c c c}   
\hline     
\hline                 
Comp & log$\xi$ &$N_\text{H}$ & Flow velocity \\
 & &($10^{21}$cm$^{-2}$) &  (km s$^{-1}$) \\
\hline                        
1 & ${2.99}\pm{0.03}$ & ${3.3}\pm{0.4}$ & ${-4260}\pm{60}$\\      
2 & ${3.70}\pm{0.04}$ & ${16.1}\pm{0.5}$ & ${-5770}\pm{30}$\\
3 & ${2.60}\pm{0.10}$ & ${1.2}\pm{0.2}$ & ${-530}\pm{10}$\\
4 & ${0.37}\pm{0.03}$ & ${0.11}\pm{0.09}$ & ${-340}\pm{10}$\\
\hline                                   
\end{tabular}
\end{table}
To complete the fit, we have also added radiative recombination continuum models where necessary and Gaussian profiles to account for visible emission lines. These narrow emission lines do not vary in flux, and as such they will not contribute to any possible delay caused by the response time of the warm absorber. The best fit model is shown in Fig. \ref{fig:rgs_spectrum}. The four distinct outflowing components were first identified by \cite{pounds2011_1}, however the parameters of each component vary slightly when compared to our results. This discrepancy is likely due to two factors. Firstly, we fit the time-averaged spectrum while \cite{pounds2011_1} have modelled the high-flux spectrum. Secondly, \cite{pounds2011_1} have calculated the physical conditions of the photoionized gas with XSTAR.\par
\subsection{Ionization and recombination rates}
\label{subsection33}
To solve the time-dependent concentrations it is necessary to know the ionization and recombination rates at each point in the light curve as well as the initial ion concentrations. Starting at the beginning of the first observation, for which we assume photoionization equilibrium, we monitor the flux level every 500s until the end of the last observation. We decided to track the flux history with this time resolution first because the typical large amplitude variability occurs on timescales of hours, and second since we would like to have a span of variability timescales that would allow us to compare with previous results on X-ray time lags, typically reaching timescales of hundreds of seconds.\par
The equilibrium ion concentrations, for each of the selected points in the light curve, were calculated with CLOUDY using the source SED and \cite{lodders2009} abundances. This is calculated for each gas component separately by feeding CLOUDY with the ionization parameter for each component, assuming equilibrium. This ionization parameter was obtained from the best fit of the time averaged RGS spectrum and then scaled to account for the flux difference, at each time bin, from the averaged flux value. If the gas responds immediately to the variability of the ionizing continuum, then the gas ionization must change according to:
\begin{equation}
 \xi=\frac{L_\text{ion}}{nr^{2}}, 
 \label{eqn:ionization_eqn}
 \end{equation}
where $\xi$ is the ionization parameter and $L_\text{ion}$ is the ionizing luminosity. The assumption here is that the distance of the gas from the ionizing source, $r$, and its density, $n$, remain constant for these timescales. Therefore, the average $\xi$ values were just linearly scaled up or down according to the flux values at each time. The ionization and recombination rates, provided by CLOUDY were stored at the end of each run. The ionization rates obtained with CLOUDY include multiple ionization after inner-shell ionization. Since in equation (\ref{eqn:time_con}) multiple ionization is not taken into account, we correct the ionization rates obtained, to force equilibrium when inserting these in the equation together with the equilibrium concentrations, which needs to yield $dn_{\text{X}^{i}}/dt=0$. 
\subsection{Time-dependent ion concentrations}
\label{subsection34}
Having all the prior information we need, we can now proceed to solve the time-dependent ion concentrations. Equation (\ref{eqn:time_con}) corresponds to a system of $N$ coupled ordinary differential equations. We first attempted to solve the system of differential equations numerically, by using a Runge-Kutta method with adaptive stepsize control, subroutine \textsc{odeint} \citep{press1992}. However our solution appeared to be \textit{stiff}, i.e. the required step size was unacceptably small compared to the smoothness of the solution for a certain time interval. To account for the stiffness of the system we have implemented in \textsc{odeint} a fourth-order Rousenbrock method, the subroutine \textsc{stiff} \citep{press1992}, which is a generalization of the Runge-Kutta method for stiff systems. \par
\begin{figure}[t]
  \centering
   \includegraphics[width=\hsize]{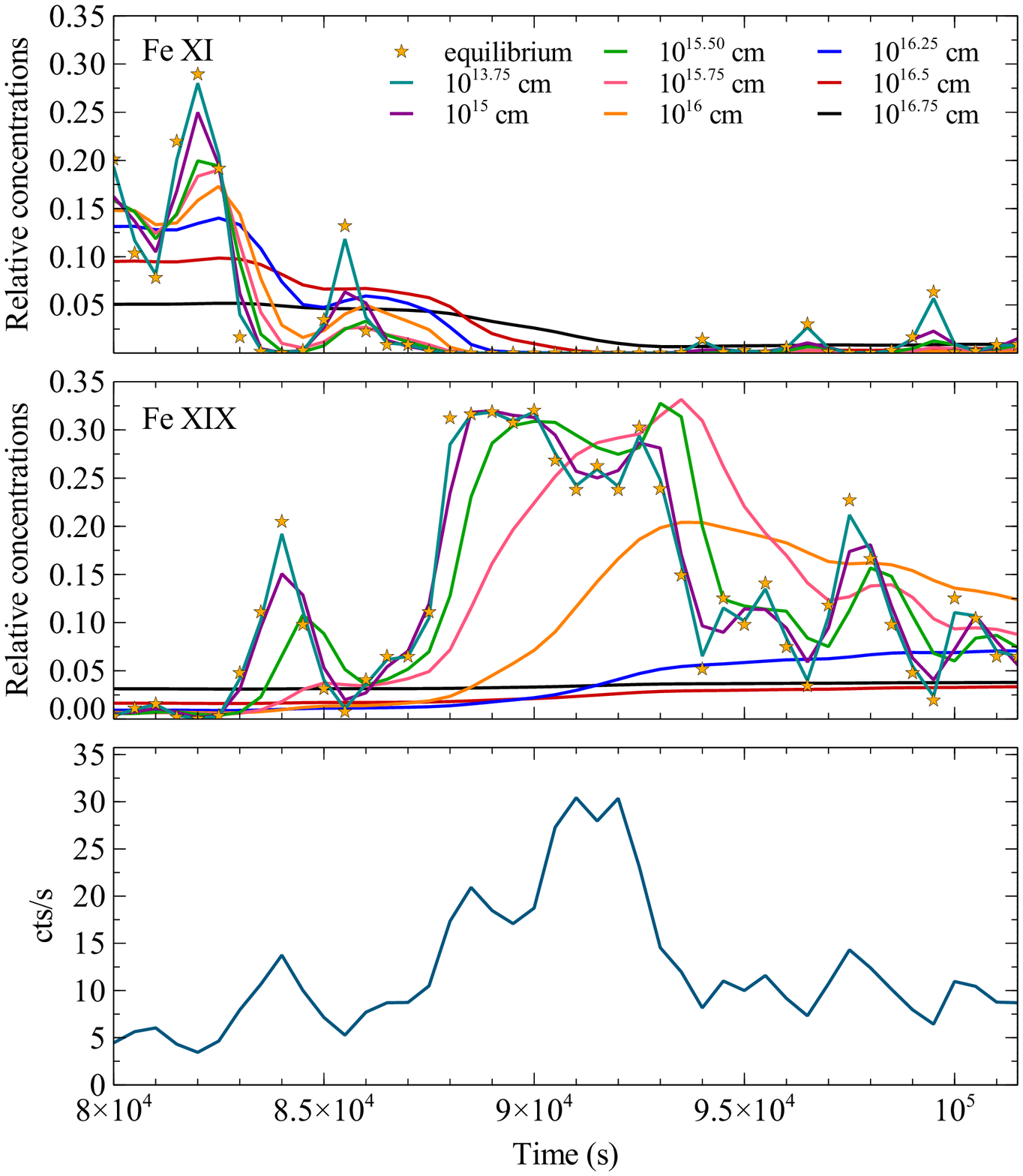}
     \caption{Illustrative example. Time-dependent evolution of the relative concentrations of \ion{Fe}{XI} (upper panel) and \ion{Fe}{XIX} (middle panel) responding to changes in the ionizing continuum (lower panel), for different distances of the gas component 1 (see table \ref{table:wa_comp}) from the ionizing source.}
      \label{fig:concentrations}
 \end{figure}

We performed the calculations for each outflow component separately, keeping the electron density, $n_\text{e}$, constant with time, as featured in \cite{kaastra2012} and \cite{nicastro1999}. As for the initial conditions, we take the first point of the first observation to be $t_{0}=0$ and feed it with the equilibrium concentrations calculated with CLOUDY. Thus, we assume equilibrium at the start of the first observation. Thereafter, for each time step the ionization and recombination rates are obtained from linearly interpolating on the ionization and recombination rates we previously calculated with CLOUDY.  These solutions are computed for a grid of distances, $r$, from the ionizing source. This allows us to explore the parameter space since the real location and density of the gas are unknown. From the observed spectrum it is possible to estimate the ionizing luminosity, $L_\text{ion}$, and the ionization parameter, $\xi$, for each outflowing component. From equation (\ref{eqn:ionization_eqn}), it follows that, for a fixed moment in time, the product of the density of the gas with its squared distance to the source, $n r^2$, is constant. Thus we assume that all the outflowing components are at a fixed distance from the central source, with a corresponding density, $n$, prescribed by the ratio $L_\text{ion}/\xi$. We then vary the distance, $r$, and calculate the time-dependent ion concentrations for the corresponding density of each component. An illustrative example of our results is shown in Fig. \ref{fig:concentrations}.
The two upper panels in Fig. \ref{fig:concentrations} display the time evolution of the relative concentration of \ion{Fe}{XI} and \ion{Fe}{XIX}, respectively, for component 1 in table \ref{table:wa_comp}. The stars represent the equilibrium concentrations, e.g. the concentrations the gas would have if it would respond instantaneously to the ionizing source variability. The coloured solid lines refer to the calculated concentrations for different distances, $r$, from the ionizing source. As a reference, for component 1, $r\sim10^{16}$ cm corresponds to a density $n\sim3.6\times10^6\ \text{cm}^{-3}$. Finally, the lower panel in the plot shows the light curve (0.3-10 keV), which is used to determine the variable ionizing flux, for the same time interval. \par
It is first interesting to note the different response of \ion{Fe}{XI} and \ion{Fe}{XIX} to changes in the ionizing flux. The outflowing component shown here has $\text{log}\ \xi\sim2.99$. At the beginning, the flux is not very high and the relative concentration of \ion{Fe}{XI} is large, while the relative concentration of \ion{Fe}{XIX} is negligible. When the flux increases significantly, the less ionized \ion{Fe}{XI} gets partially stripped in favour of its more ionized relative \ion{Fe}{XIX}. This is exactly the behaviour we expect, since a significant increase in flux will photoionize Fe and take it to its higher levels of ionization. Likewise, a significant decrease in flux will lead Fe to recombine and we will see a higher concentration of the less ionized levels. We use Fe as an example since it is sensitive to ionization parameters such as this one. Also worth noting in Fig. \ref{fig:concentrations}, is the detection of the three different kinds of possible behaviour for the response of the gas relative to the ionizing continuum (see introduction to section \ref{section3}). For close distances to the ionizing source, the response is almost instantaneous. This is the case for distances $\le 10^{14} \text{cm}$, in which the response is immediate but the ion concentration may not have time to reach the equilibrium values before being disturbed again.  For distances between $\sim10^{15} \text{cm}$ and $\sim10^{16.5}\text{cm}$, the gas responds to the variations of the ionizing continuum with a visible time delay. This is the most interesting case for our study, since it is the case for which a time lag associated with the response time of the gas relative to the continuum variability may be detected. Finally, at distances $\ge 10^{16.5} \text{cm}$, the gas is unable to respond to the changes of the continuum flux. For distances so far from the central source, the gas is in a steady state and the relative ion concentrations vary slightly around a mean value, corresponding to the mean flux level over time. We explore in section \ref{section4} which effects the complex behaviour of the warm absorber can have on the X-ray time lags.
\subsection{Simulated spectrum and light curve generation}
\label{subsection35}
 \begin{figure}[b!]
  \centering
   \includegraphics[width=\hsize]{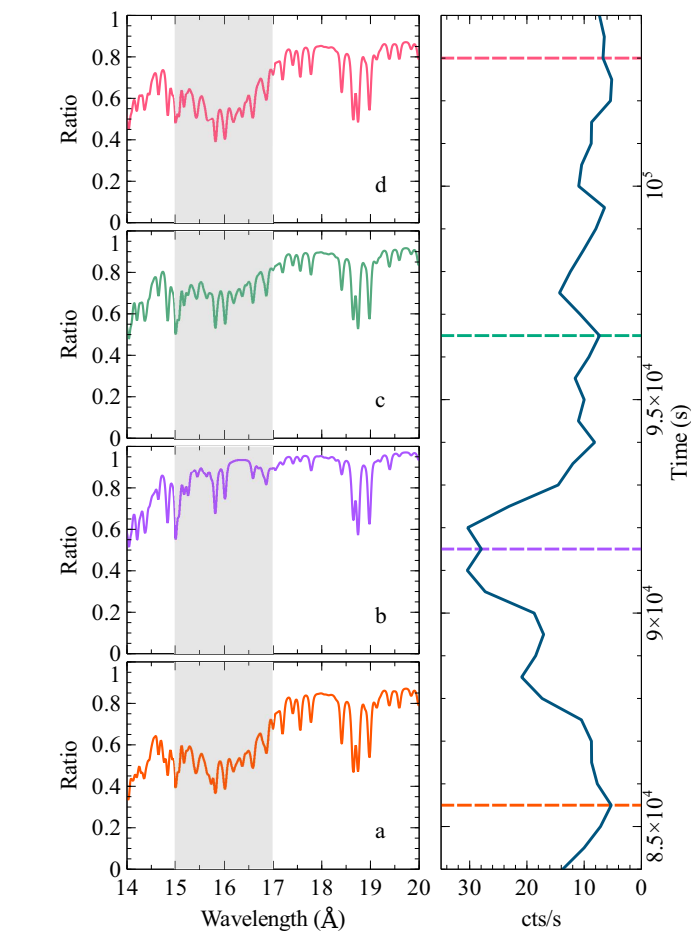}
     \caption{Illustrative example: The left panel shows the data/continuum model ratio of the simulated RGS spectra for $r=10^{15.75}$cm, around the region of the Fe UTA complex (15-17$\ \AA$, grey area in the spectra), for distinct time steps, which are indicated by the dashed lines on the right. The right panel shows a slice of the generated light curves from the simulated spectra, that includes the time steps of the spectra in the left. Note that spectra (a), (c) and (d) have approximately the same flux level but show different absorption features, indicating the dependence of the warm absorber on the flux variation history.}
      \label{fig:rgs_simulated}
 \end{figure}
In order to study the pure effects of a variable warm absorber on the X-ray time lags of NGC 4051, it is necessary to generate light curves which are only affected by the changes of the relative ion concentrations over time. We then proceeded to the simulation of RGS and EPIC-pn spectra which we later use to build the necessary light curves for our analysis. The aim is to construct new light curves with the same flux history as those of NGC 4051, but whose spectrum at each time step is only affected by the variable warm absorber. \par
Using SPEX, we simulate RGS and EPIC-pn spectra for each time step in the following manner. We included the predefined continuum shape (see sections \ref{subsection31}, \ref{subsection32}) with the flux levels prescribed by the light curve. The shape of the spectrum is kept constant with time to avoid time delays caused by intrinsic changes in the X-ray continuum shape. To account for the variability in the ionizing source, we scale the flux level of the X-ray spectrum according to the count rate history of the X-ray light curves of NGC 4051. In fact, the spectral shape of NGC 4051 is highly variable with time. However this should not invalidate our results for the following reasons. First because the absorption effects are stronger in the soft X-rays whereas the spectral-variability more strongly affects the shape of the continuum above $\sim1$ keV \citep[see e.g.][]{uttley99,vaughan11}. Second, since the source count rate is dominated by the soft photons, the soft X-rays are a good proxy for the ionizing flux changes. This continuum is absorbed by a multi-layer warm absorber, with ionic ion column densities given by the time-dependent ion concentrations. This is achieved using the \textsc{slab} model from SPEX. The \textsc{slab} model is a thin slab absorption model, for which the ion column density can be chosen independently. The column densities are calculated, for each ion, with the time-dependent concentrations we computed previously. In this way each simulated spectrum is identical in underlying continuum shape, and only changed by the variable column densities derived from the time-dependent concentrations. The complex warm absorber we find in the RGS data is also added to the EPIC-pn continuum, since it must also be present there, only we do not have the spectral resolution to observe the narrow absorption features. The model is then folded through the instrumental response matrix to obtain the net count rate per energy channel. Since our purpose for now is to study the unadulterated effect of the variable warm absorber in the absence of observational noise, no Poisson noise was added to the simulations (for implications of noise for lag detectability see section \ref{sectiondetectability}). Examples of the simulated spectra are shown in Fig. \ref{fig:rgs_simulated} (left panel).
The simulated spectra are subsequently used to generate light curves for all bands of interest, both for EPIC-pn and RGS. An example of the generated light curves is also shown in Fig. \ref{fig:rgs_simulated} (right panel). Furthermore, in Fig. \ref{fig:rgs_simulated} we observe the delayed response of the gas to the variations of the continuum. The figure shows the simulated spectra resulting from the calculated concentrations at $r=10^{15.75}$ cm. As we can see from Fig. \ref{fig:concentrations}, the ion concentrations follow the light curve evolution with a significant time delay. This time delay is also detectable in a simple example in Fig. \ref{fig:rgs_simulated}. From the time step represented in spectrum (a) to the time step represented in spectrum (b), the flux increases by a factor of $\sim 3$. The changes in the spectral features are visible. When the flux decreases to the same level as before, in the  time step represented in spectrum (c), the spectrum has not yet recovered the same features it had at the beginning. Some time later, at the time step represented in spectrum 'd', the spectrum finally starts to recover its initial shape for the same flux level. This is a signature of the time delay observed in Fig. \ref{fig:concentrations}, and, in this case, is associated with a recombination timescale on the order of at least hours.\par
Those small changes in absorption are indeed responsible for any time delays, since no other variability processes are considered here. The absorption changes are small in general and will modulate only a small fraction of the total flux. These small modulations will introduce small variations in the distribution of flux in the light curves, which can then allow us to measure time delays between the absorbed bands and the continuum. In the next section we make use of the light curves generated with the simulated spectra and compute the spectral-timing products resulting from the application of Fourier methods.
%
\section{Timing analysis}
 \label{section4}
In the previous section we have shown how we generate light curves which contain only the flux variability plus the effects of the variations in the relative ion concentrations due to the intrinsic source variability, excluding the effects associated with variability of the continuum spectral shape. In this section, we analyse these time series in the Fourier domain and assess the effects of the variable warm absorber on the lag-frequency and lag-energy spectra of NGC 4051. In the next paragraphs we introduce the Fourier techniques commonly used to detect time-lags in X-ray light curves. This is intended to give an introduction to timing analysis for readers that are not familiar with the subject. For more details about spectral-timing analysis on X-ray light curves of black hole systems of all scales see \cite{uttley14}.\par
The most common tool in Fourier analysis is the Fourier power spectrum, or power spectral density function (PSD). The power spectrum contains information about the underlying structure of a stochastic variability process. It shows the dependence of the variability amplitude as a function of temporal frequency. However, we cannot simply measure the power spectrum. Since every observation is a random realization of the underlying PSD, what we actually measure is a noisy random realization, referred to as the periodogram. The periodogram is obtained from the discrete Fourier transform of the signal we are analyzing. Thus, we begin by introducing the discrete Fourier transform $X_{n}$ of a time series $x$, 
\begin{equation}
X_{n}=\sum\limits_{k=0}^{N-1} x_k \text{exp} (2\pi i n k/N) ,
\end{equation}
where $N$ is the number of time bins of width $\Delta t$ and $x_k$ is the value of the light curve at index $k$. The Fourier transform is evaluated at each Fourier frequency $f_n=n/(N\Delta t)$, with $n=1,2,...,N/2$. Thus, the lowest Fourier frequency is $1/N\Delta t$, which corresponds to the inverse of the total observation time, and the highest Fourier frequency is $1/2\Delta t$, also known as the Nyquist frequency. The periodogram is defined as the modulus square of the Fourier transform of the time series $x$, $| X_{n} |^{2}$, which can be written as
\begin{equation}
| X_{n} |^{2}=X_{n}^{\ast} X_{n}
\end{equation}  
where the $X_{n}^{\ast}$ represent the complex conjugate of $X_{n}$. The periodogram can be normalized to yield the same units as the PSD and is usually computed for several light curve segments. It is subsequently binned over the segments, and also over frequency bins, to obtain an estimate of the underlying PSD, which can then be modelled. Following the notation from \cite{uttley14}, we denote the binned periodogram by $\overline{P}_{X}$.\par
We now consider two light curves, x and y, with x being a band of interest (typically a small energy bin), and y a broad reference band that excludes x. It is possible to derive the frequency-dependent lag between these two bands by applying the cross-spectrum method. The cross-spectrum is defined as
\begin{equation}
C_{XY,n} = X_n^{\ast}Y_n
\label{eq.2}
\end{equation}
 and when considering the complex polar representation of the Fourier transforms, equation (\ref{eq.2}) takes the form
 \begin{equation}
 C_{XY,n} = A_{X,n} A_{Y,n} \text{exp}(i\phi_n ),
 \end{equation}
where $A_{X,n}$ and $A_{Y,n}$ are the amplitudes of the Fourier transform and $\phi_n$ the phase lag between the two bands. In order to reduce noise, the cross-spectrum is also computed for several light curve segments and averaged over segments and over frequency bins. The averaged cross-spectrum is denoted as $\overline{C}_{XY,n}$. For an easier interpretation, the phase lag can be converted into a time lag
\begin{equation}
 \tau(\nu_j)=\phi(\nu_j)/(2\pi\nu_j),
 \end{equation}
 from which we directly infer the time delay between the two time series. The error on the lag is dependent on the coherence. The coherence measures the level of linear correlation between two signals, which in our case corresponds to two light curves of different energy bands. The coherence is defined as
\begin{equation}
\gamma^2(\nu_j)=\frac{\left| \overline{C}_{XY}(\nu_j) \right|^2-n^2}{\overline{P}_X(\nu_j) \overline{P}_Y(\nu_j)}
\end{equation} 
and takes values between 1 (for perfectly correlated signals) and 0 (for signals with no linear correlation). The $n^2$ is a bias term due to Poisson-noise contribution to the modulus square of the cross-spectrum \citep[see for further details][]{uttley14,vaughan1997}. The error on the phase lag is then defined as
\begin{equation}
\Delta\phi(\nu_j)=\sqrt{\frac{1-\gamma^2(\nu_j)}{2\gamma^2(\nu_j)KM}}
\label{eqn:delta_phi}
\end{equation}
where $K$ and $M$ respectively correspond to the number of frequencies and segments over which the periodogram and the cross-spectrum were averaged. Converted to the time domain, equation (\ref{eqn:delta_phi}) takes the form
\begin{equation}
\Delta\tau=\Delta\phi/(2\pi\nu_j)
\end{equation}
\par
Through the application of these Fourier techniques, it is possible to detect the timing properties of the time series, for a broad range of timescales. We applied these methods to the light curves we have generated from the simulated spectra, which account for the warm absorber response to changes in the ionizing continuum. Our results are presented below.
\subsection{Lag-frequency spectrum}
\begin{figure}[b!]
   \centering
   \includegraphics[width=\hsize]{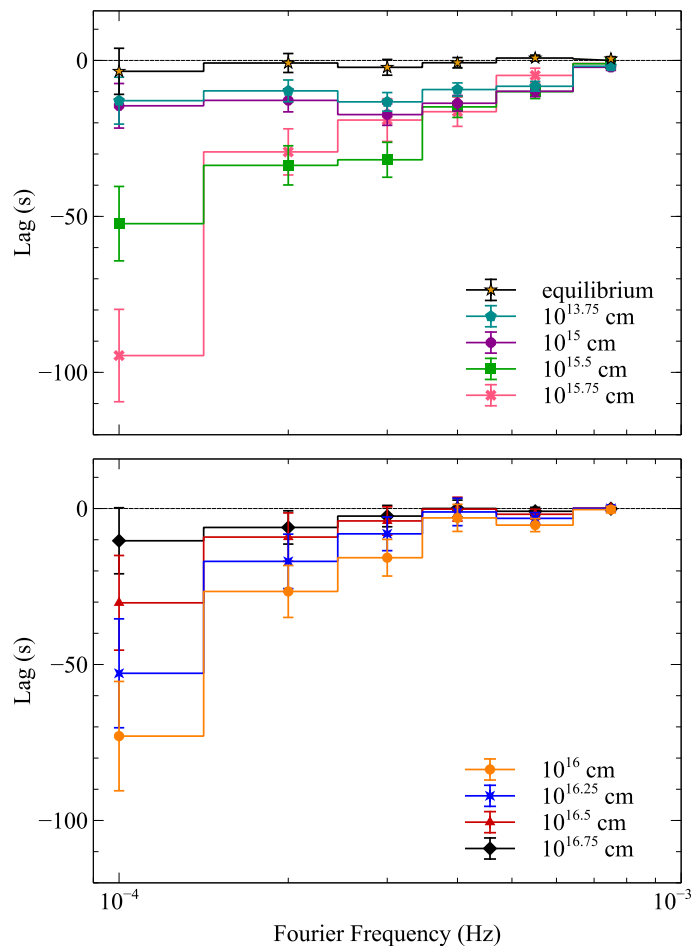}
     \caption{Lag-frequency spectrum for a grid of distances of the warm absorber to the central source. A negative lag indicates that the soft band (0.3 - 1.0 keV) lags the hard (2.0 - 5.0 keV).}
       \label{fig:lag_frequency_several}
  \end{figure}
The lag-frequency spectrum is a common way to represent the time (or phase) lag between two time series as a function of Fourier frequency. Mostly for AGN, this way of representing the timing properties of the light curves, displays a very characteristic pattern, consisting of two distinct signatures \citep[e.g.][]{zoghbi2011,demarco2013}. At low frequencies, which correspond to long timescales, the hard photons from a given variation lag the soft photons from that same variation. This delay is thus called a hard lag, and can have a magnitude from tens to thousands of seconds. At higher frequencies, the sign of the lag is inverted. Therefore, on short timescales, the soft photons from a certain variation arrive after the hard photons from that same variation, and this is called a soft lag. This behaviour is often observed when selecting a soft (e.g. 0.3-1.0 keV) and hard (e.g. 2.0-5.0 keV) time series, and computing the cross-spectrum. Since the warm absorber greatly affects the soft X-rays, we investigate how the soft X-rays behave compared to the hard X-rays. \par
We use an EPIC-pn simulated time series in the soft band (0.3-1.0 keV) and compute its cross-spectrum with a harder band (2.0-5.0 keV). To compute the cross-spectrum we divided the data into contiguous segments of 10 ks each and a time bin of 500s, the same time resolution we have for the time dependent concentrations. We do this for the same range of distances, of the warm absorber to the central source, that we used to compute the time-dependent concentrations. Our results are presented in Fig. \ref{fig:lag_frequency_several}.
The upper panel shows the lag vs frequency spectra for the equilibrium situation and for distances from $10^{13.75}$ up to $10^{15.75}$ cm. The lower panel shows the same but for distances from $10^{16}$ up to $10^{16.75}$ cm. For an equilibrium situation, represented by the stars in Fig. \ref{fig:concentrations}, the warm absorber responds instantaneously to changes in the ionizing continuum. Since the warm absorber has an instantaneous response to the source variability, we do not expect to observe a time delay between the more absorbed soft X-rays and the hard X-rays. This is the case for warm absorbers located at smaller distances from the black hole. As seen in Fig. \ref{fig:concentrations}, for radial distances up to $10^{14}$ cm, the warm absorber appears to be in equilibrium with the variability of the central source, responding instantaneously. For distances $\ge10^{15.75}$ cm we observe in Fig. \ref{fig:concentrations} that the warm absorber is no longer in equilibrium with the ionizing continuum. This non-equilibrium situation is characterized by a response time, which causes a delay between the more absorbed photons and the continuum photons. This delay is detected in the lag-frequency spectrum we computed. The time lag between the soft and the hard photons is detected towards lower Fourier frequencies, indicating long timescales. The strongest delay appears to come from the region around $10^{15.75}$ cm. For even larger radial distances, the delay between the soft and the hard bands smooths away, which again agrees with what we expect, since for larger distances the warm absorber appears to be in a steady state, possibly in equilibrium with the average luminosity over time, and is not able to respond to the fast variability of the source. Hence, we proceed to analyse the most interesting situation for our work, which corresponds to the distance capable of producing the strongest time delay overall, $r=10^{15.75}$ cm. As a reference, for component 1, $r\sim10^{15.75}$ cm corresponds to a density $n\sim1.1\times10^7 \text{cm}^{-3}$. The corresponding densities for the other components, at these distances, may be easily computed as $n_{\text{comp}_\text{Y}}=\xi_{\text{comp}_\text{X}} n_{\text{comp}_{\text{X}} }/\xi_{\text{comp}_\text{Y}}$, where X and Y can be any of the components in table \ref{table:wa_comp}. \par
\begin{figure}[t!]
   \centering
   \includegraphics[width=\hsize]{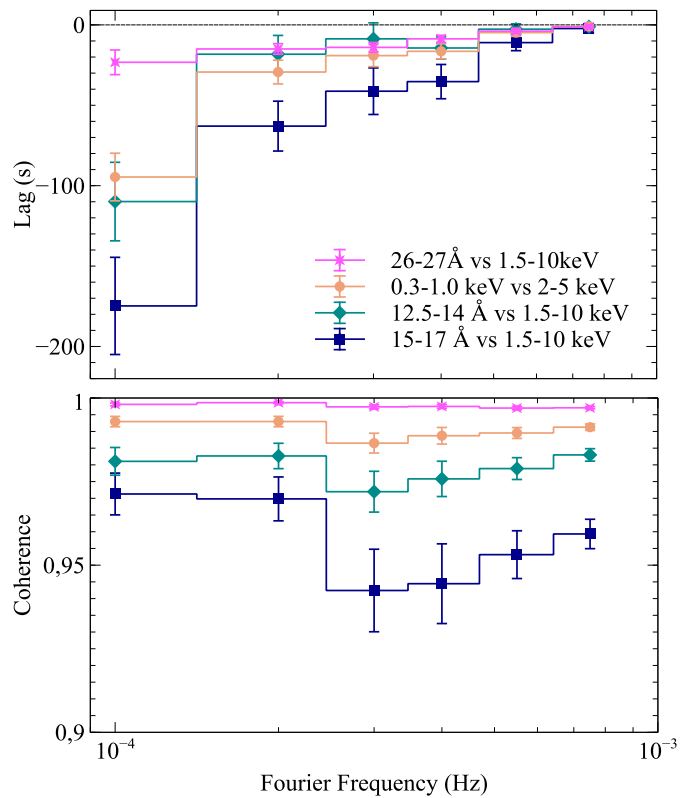}
     \caption{Lag-frequency spectrum (top) and coherence spectrum (bottom) of several bands of interest for $r=10^{15.75}$cm. Label reads from top to bottom: unabsorbed band RGS (26-27 $\AA$) vs broad band EPIC-pn (1.5-10 keV), soft band (0.3-1 keV) vs hard band (2-5 keV) both from EPIC-pn, Ne blended absorption in the RGS band (12.5-14 $\AA$) vs broad band EPIC-pn (1.5-10 keV), and Fe UTA absorption feature in the RGS band (15-17 $\AA$) vs broad band EPIC-pn (1.5-10 keV). A negative lag indicates that the band of interest lags the reference band.}
       \label{fig:lag_frequency}
  \end{figure}

For the purpose of studying this situation in detail, we compute the cross-spectrum again, this time using narrow energy bins of interest. The energy bins are selected by taking into account the absorption features observed in the RGS spectrum of NGC 4051. An ideal feature to study the time behaviour of the warm absorber is the Fe UTA complex (15-17 $\AA$, see Fig. \ref{fig:concentrations}). This feature shows high opacity and it is very sensitive to flux variations \citep[see e.g.][]{krongold2007}. Furthermore, the Fe UTA complex is broader compared to a single absorption line, and so it also provides us with higher signal to noise. Taking a time series from the simulated RGS light curves in this band, we compute its cross-spectrum with a broad EPIC-pn simulated light curve. For comparison, we also computed the cross-spectrum for another absorbed band, 12.5-14 $\AA$, and for a band which shows only weak absorption, 26-27 $\AA$. These results are displayed in Fig. \ref{fig:lag_frequency}, together with the lag-frequency and coherence spectra for the hard and soft bands.\par
The time delay between the Fe UTA feature and the broad continuum band is $\sim175$ s, at a Fourier frequency of $10^{-4}$ Hz. In contrast, the time delay between the band that shows almost no absorption and the broadband continuum is only $\sim25$s, for the same frequency. Furthermore, the time delay between the absorbed 12.5-14 $\AA$ band and the continuum is $\sim110$s. Even using EPIC-pn data, which are not as sensitive to the changes in these features, we observe a soft lag of $\sim95$s at low frequencies. From these results alone we can conclude that a warm absorber with the characteristics of the one in NGC 4051, if located at a certain, optimal distance from the central source ($10^{15.75}$ cm with our assumptions), would produce a time delay of $\sim100$s between the more absorbed bands and the broad ionizing continuum, on timescales of hours or longer. This lag however is not to be interpreted directly as a response time. The lags are effectively diluted since only a small fraction of the total flux is being modulated by the variations of the absorbing features \citep[for a discussion on lag dilution see e.g.][]{uttley14}.\par
As can been seen in the lower panel of Fig. \ref{fig:lag_frequency}, the coherence between the two energy bands also appears to show a small dependence on the absorption. For increasing lag, the coherence between the two bands is slightly reduced. However, the whole process still appears highly coherent, reaching a coherence of $\sim 0.97$ for the longest hard lag detected. Interestingly, the observed coherence at low frequencies for NGC 4051 is high, but not exactly 1. In fact the observed coherence for NGC 4051 at low frequencies, has a value of $\sim 0.85$ \citep{alston2013}. This could be, in part, due to the small effect of the warm absorber.\par
To further investigate the timing behaviour, for several narrow energy bins covering the whole RGS spectral range, we have proceeded to compute the lag-energy spectrum.
\subsection{Lag-energy spectrum}
The energy resolved lag spectrum, or lag-energy spectrum \citep{uttley14}, is an extremely valuable tool for our study, since we can relate it directly to the observed energy spectrum and assess the behaviour of the lags at energies that are highly absorbed. Hence, we are able to directly link the lags to the regions in the spectrum where absorption plays a major role.  Taking the cross-spectrum between each one of the channels of interest versus the reference band, one can select a range of Fourier frequencies of interest and take its average, plotting then the average lag against energy. In this way we can build a lag-energy spectrum which contains the information of all the previously calculated lag-frequency spectra, but then taking into account a specific range of frequencies. This way of looking at time lags allows to compare the relative lag in several bands versus the reference band, providing a direct comparison of the behaviour of the time lags and the energy spectrum, since both are a function of energy. \par
  \begin{figure}[b!]
  \centering
   \includegraphics[width=\hsize]{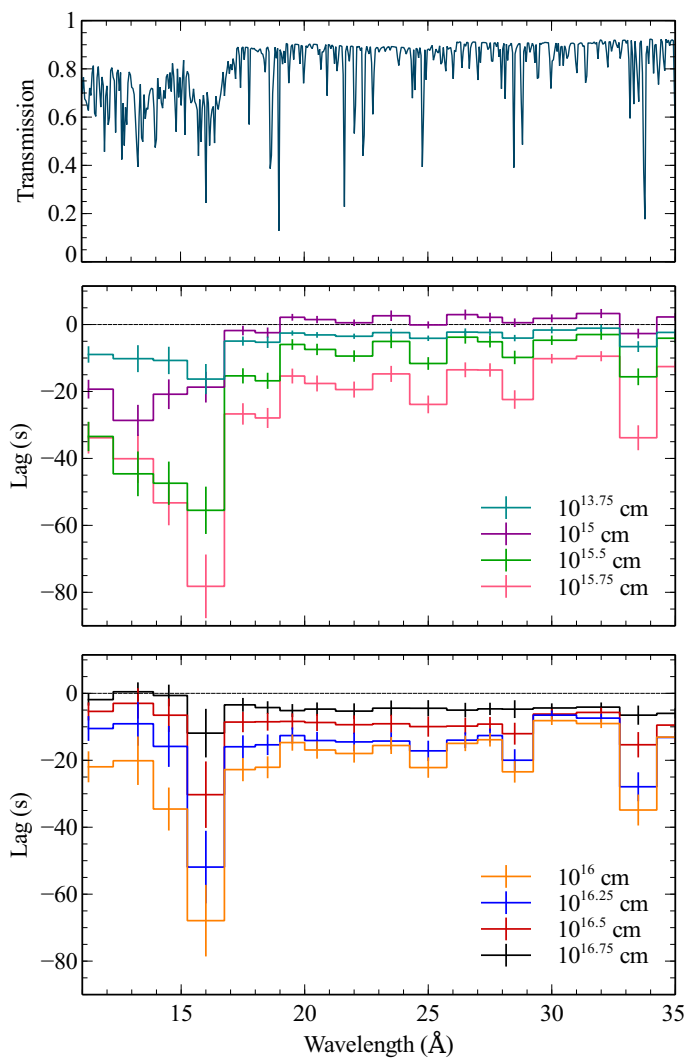}
     \caption{Warm absorber transmission model (top) and lag-energy spectrum (middle and bottom) for the computed grid of distances. The reference band to compute the lag-energy spectrum is a broad EPIC-pn band (1.5-10 keV). Note that a negative lag indicates a soft lag, which means that the band of interest lags the broad reference band. This helps to easily match the lag-energy spectrum to the absorption features in the energy spectrum.}
      \label{fig:lag_energy_several}
 \end{figure}
Taking into account the fact that any time delays between the more absorbed bands and the continuum, appear to be negligible above $3-4\times10^{-4}$ Hz, we have considered the range of frequencies $1-3\times10^{-4}$ Hz to compute the lag-energy spectrum. In Fig. \ref{fig:lag_energy_several}, we plot the transmission of the four warm absorber models, contained in our best fit to the averaged RGS spectrum, and the lag-energy spectrum for the computed grid of distances.
 As can be seen from Fig. \ref{fig:lag_energy_several}, the regions in the energy spectra from where more direct continuum is removed due to absorption from the outflowing gas, correspond to the region of the lag-energy spectra where we see a stronger soft lag. The amplitude of the lags results from averaging in Fourier frequency and is therefore lower than the maximum lag observed in the lag-frequency spectrum. The lag appears to be maximal around distances of $10^{15.75}$ cm, for the whole energy spectrum.\par
\begin{figure}[t]
  \centering
   \includegraphics[width=\hsize]{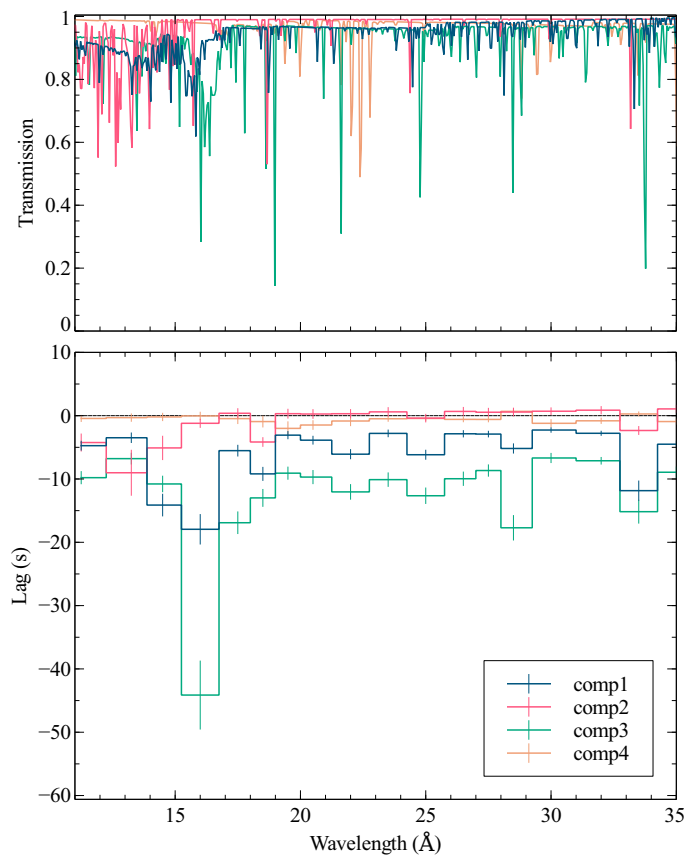}
     \caption{Warm absorber transmission model and lag-energy spectrum for each warm absorber component separately, for $r=10^{15.75}$ cm. See table \ref{table:wa_comp} for the parameters of the gas components. Note that a negative lag indicates a soft lag, which means that the band of interest lags the broad reference band. This helps to easily match the lag-energy spectrum to the absorption features in the energy spectrum.}
      \label{fig:lag_energy_comp}
 \end{figure}

To assess the contribution of each warm absorber component to the lag-energy spectrum, we have re-done the spectral simulations and cross-spectral products for each warm absorber component separately. This was once again done for the most interesting case, corresponding to a radial distance of $10^{15.75}$ cm, where a maximal lag is seen for the whole energy range. The results are shown in Fig. \ref{fig:lag_energy_comp}.
 When analysing the lag-energy spectrum of each component separately, we see that the warm absorber which has a higher impact on the lags, in this source, is the one associated with the Fe UTA feature, since this is the most prominent feature in the whole energy spectrum, and is also very sensitive to the flux variations (see Fig. \ref{fig:rgs_simulated}). This corresponds to component 3, which has a mild ionization parameter of $\text{log}\xi\sim2.6$. Component 1, has an ionization parameter close to the one of component 3, $\text{log}\xi\sim2.9$, and so it affects the same energy bands as component 3, contributing to a stronger effect at those energies. Component 2 is the component with the highest ionization parameter, $\text{log}\xi\sim3.7$, and so it is the major contributor at high energies (around 10-15 $\AA$). Finally the less ionized component, component 4, does not appear to have a major contribution. This can in part be explained by the low ionization of this component. At the same distance, component 4 will be much denser than the other components, and a higher density would result in a faster response time. However, since the column density of this component is very low when compared to the others, its contribution to absorption of the spectrum (see the transmission plot from Fig. \ref{fig:lag_energy_comp}), and therefore to the lags is minimal even at the larger distances and/or lower densities.\par
Note that the response time is not only dependent on the density of gas, but it also depends on the ion species that play a stronger role for each component (see equation \ref{eqn:time_con}). Therefore, the contribution to the lags only scales linearly with density, if the same warm absorber component is being considered.
%
\section{Comparison to the real data}
 \label{section5}
 Having studied the pure effects that a variable warm absorber would introduce to the timing properties of the X-ray light curves of NGC 4051, we now compare our results with the observed data and discuss whereas such effects could be detectable with currently available instruments and data. 
\subsection{Detectability}
\label{sectiondetectability}
\begin{figure}[b!]
  \centering
   \includegraphics[width=\hsize]{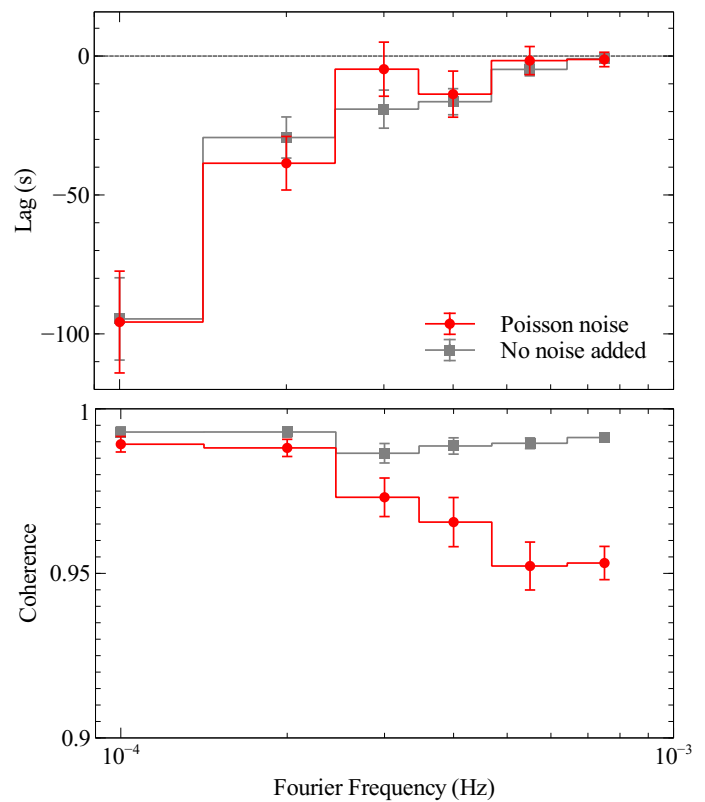}
     \caption{Lag-frequency spectrum (top panel)  and coherence spectrum (lower panel) between the soft (0.3-1.0 keV) and hard (2.0-5.0 keV) bands of the simulated light curves for EPIC-pn including Poisson noise.}
      \label{fig:lag_frequency_detectability}
 \end{figure}

To assess the detectability level of these warm absorber effects we add Poisson noise when generating light curves from the simulated spectra. We perform this analysis for the soft (0.3-1.0 keV) and hard (2.0-5.0 keV) bands and compute the lag-frequency spectrum using the methods outlined in section \ref{section4}. The results are shown in Fig. \ref{fig:lag_frequency_detectability}.\par
As can be seen from the lag-frequency spectrum, Poisson noise does not play a major role at the Fourier frequencies of interest. It minimally increases the error bars on the computed lags and scatters the value of the lag slightly. This is to be expected since from the study of the power-spectrum of NGC 4051, it is possible to observe that the Poisson noise only has an effect at high Fourier frequencies \citep{alston2013}. Therefore, the effects on the lags are minimal. As for the coherence, it is also only affected towards higher Fourier frequencies. In this way, it is safe to say that the effects we discussed in the previous section are detectable using currently available data. This means that if the warm absorber in NGC 4051 was located at a radial distance of $\sim 10^{15.75}$cm from the black hole, and if no other processes were at play, a soft lag would be expected at low frequencies. 
\subsection{Comparison to RGS data}
\begin{figure}[t]
  \centering
   \includegraphics[width=\hsize]{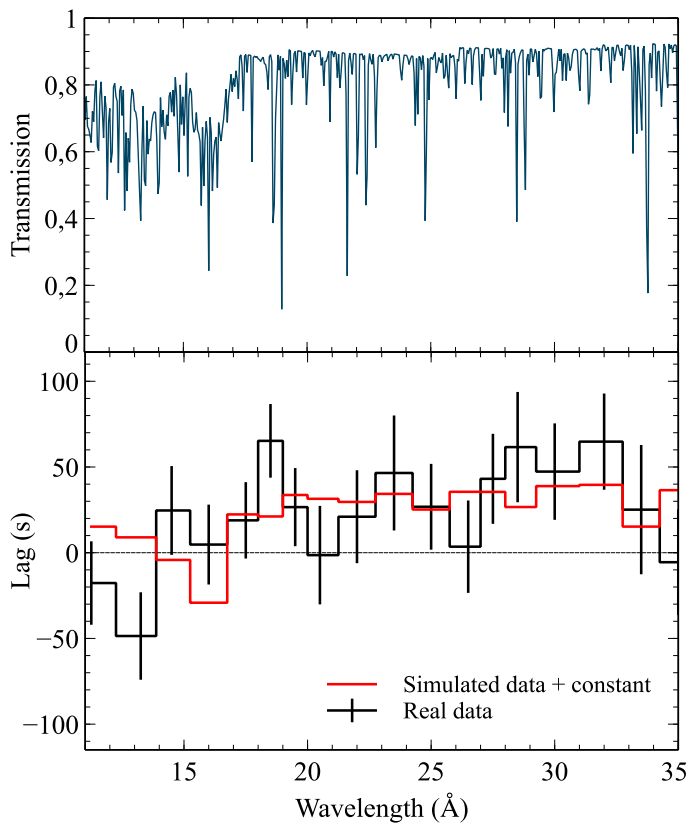}
     \caption{Warm absorber transmission model (top) and lag-energy spectrum (bottom). The lag-energy spectrum from simulated data, at $r=10^{15.75}$cm (see Fig. \ref{fig:lag_energy_several}), with an additive constant of 49.1 s, was fitted to the observed lag-energy spectrum, plotted in black, to obtain a $\chi^{2}$ of 20.7 for 17 d.o.f.. The simulated lag-energy spectrum with an offset, corresponding to the constant that resulted from the fit ($\sim 50$ s), is overplotted in red. The reference band to compute the lag-energy spectrum is a broad EPIC-pn band (1.5-10 keV). Note that a negative lag indicates a soft lag, which means that the band of interest lags the broad reference band. This helps to easily match the lag-energy spectrum to the absorption features in the energy spectrum.}
      \label{fig:rgs_real_lag_energy}
 \end{figure}

We now compare the results we obtained from the simulated RGS data to the real observations. The lag-energy spectra that we presented in Fig. \ref{fig:lag_energy_several} and Fig. \ref{fig:lag_energy_comp}, were computed from simulated RGS data that include only the effects of the response of the warm absorber to the changes in flux of the ionizing continuum. The RGS spectra, due to the higher spectral resolution of the RGS instrument, helps us to distinguish what are the spectral features that have a higher contribution to the lags. We have then computed the lag-energy spectrum using real RGS light curves, in bands of interest, to investigate whether it would be possible to detect, in the real data, a signature of the warm absorber, even in the presence of external continuum processes, that we did not consider in the simulations. We find that for the real data, the size of the error bars makes it difficult to say if the lag is associated with the warm absorber or if the continuum processes dominate. To examine this further we have performed a simple test. We consider that the photons that reach the warm absorber have already been subjected to the continuum processes that produce the dominant observed hard lag at low Fourier frequencies, and thus already carry a time delay associated with these processes. When these photons reach the warm absorber and are subjected to the response time, this effect should be additive. Thus, the lag caused by response time of the warm absorber should add to the previously carried lag. Hence, our simplified model consists in fitting the simulated lag-energy spectrum with an additive constant to the observed lag-energy spectrum. We found that adding a constant lag of $\sim50$ s provides the best fit from matching our simplified model to the data. These results are shown in Fig. \ref{fig:rgs_real_lag_energy}. Therefore, we can conclude that it is not possible, with this dataset, to detect structure in the lag-energy spectrum that may indicate a signature of the response time of the warm absorber to changes in the ionizing continuum. Other datasets, and in particular future instrumentation with higher quality grating or calorimeter data may allow this test.\par
 \subsection{Effects on measured X-ray time lags}
 Understanding and disentangling the various components that play a role in the timing properties of the X-ray light curves of AGN is fundamental to break the degeneracies that spectral fitting alone perpetuates. It is then crucial to determine the level to which the effects of the variable warm absorber may affect the standard picture of the lag-frequency spectrum of AGN X-ray time series.\par
Comparing to the previous work on the timing properties of NGC 4051 by \cite{alston2013}, it is clear that a soft (negative) lag is not detected at low frequencies when using the whole 2009 dataset. We have performed the same analysis on the real data and our results agree with \cite{alston2013}. The lag-frequency spectrum of NGC 4051, when considering the whole 2009 dataset, consists of a hard (positive) lag at low frequencies followed by a turnover that leads to a soft (negative) lag at higher frequencies. As mentioned earlier on this paper, the soft lag at high frequencies has been connected to a direct result of reflected emission of the hard coronal X-rays on the softer disc photons, causing a time delay between the direct hard X-ray emission and the reflected one \citep{fabian2009}. The variable warm absorber also causes a delay between soft and hard photons, however only for long timescale variations (on the order of hours). Thus we do not expect the measured reverberation lags, in this source, to be affected by the variable warm absorber. However, the soft lag we detect at low frequencies (long timescales) due to the delayed response of the more absorbed energy bands compared to the ionizing continuum, can affect the measured hard lag at those frequencies. This could in principle mean that the intrinsic hard lag at low frequencies would have a higher amplitude if we would be able to exclude the effect from the warm absorber.\par 
Hard lags at low Fourier frequencies are common in AGN lag-frequency spectra, specially for systems with a relatively low mass such as NGC 4051 \citep{demarco2013}. Therefore, we stress that understanding the effect of the warm absorber is necessary in order to fully grasp the nature of the hard lags in these sources. 
 \subsection{Flux dependence}
 The shape of the SED and the variability amplitude of NGC 4051 are known to vary with flux level \citep{vaughan2001}. To investigate the effects of these variations with flux level, \cite{alston2013} performed a flux resolved study of the lag-frequency spectrum of NGC 4051. \cite{alston2013} found that the lag-frequency spectrum of NGC 4051 is also variable. When considering only high flux segments, the lag-frequency spectrum resembles the one calculated with the whole time series. Such behaviour is expected since the high flux segments are also the most variable and will dominate the averaged cross-spectrum due to the higher Fourier amplitudes. On the other hand, the lag-frequency spectrum selecting only medium or low flux segments shows a transition at low Fourier frequencies (long timescales); a hard lag is no longer detected but instead a soft lag is observed. It has not yet been possible to physically explain such behaviour. However, since the low flux lag-frequency spectrum resembles the soft lags produced by a warm absorber, we also performed this analysis on our simulated data. \par
\begin{figure}[t]
  \centering
   \includegraphics[width=\hsize]{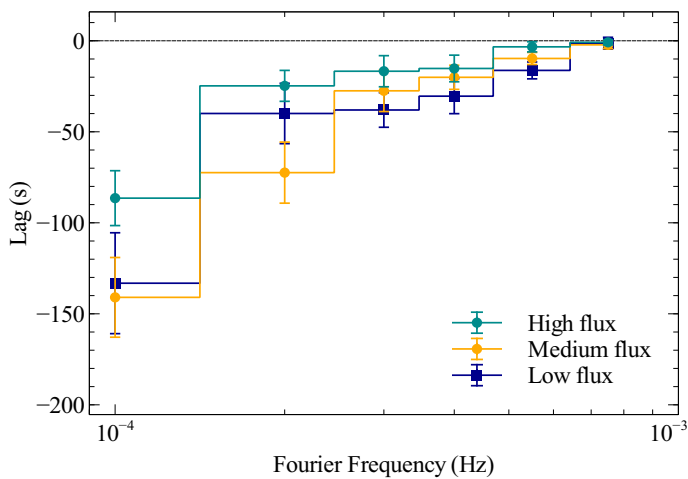}
     \caption{Lag-frequency spectrum between the soft (0.3-1.0 keV) and hard (2.0-5.0 keV) bands of the simulated light curves for EPIC-pn, for different flux levels.}
      \label{fig:flux_levels}
 \end{figure}

Using the same reference mean flux levels as \cite{alston2013}, we compute the flux resolved lag-frequency spectrum to find whether the soft lag we find due to the variable warm absorber could also be dependent on the flux level. We over-plot in Fig. \ref{fig:flux_levels} the lag-frequency spectrum for the high, medium and low flux level segments. We find that the soft lag we observe exclusively due to the variable warm absorber appears to be more significant when picking only low flux segments. The lag depends on delayed changes in the fraction of absorbed photons. When this fraction is large, e.g. at low fluxes, the lag is large, whereas when it is small, e.g. high fluxes, the lag is small. From the simulated spectra, we have verified that these flux-dependent changes in the absorbed fraction of photons do occur for some absorption features, particularly for the Fe UTA feature, which is one of the main contributors to the lag. Additionally, different atomic transitions are affected by different ionization states, which result in different response times and responses at different timescales. We suggest that the flux dependence of the lags at low Fourier frequencies may be at least in part due to ionization state of the gas, but investigating in detail this complex behaviour goes beyond the scope of this paper and we will address it in future work.
As for the non-detection of a soft lag at low frequencies in the total/combined dataset, this could be due to the fact that the high flux states are dominated by other processes. Possibly the mechanism responsible for the hard lags at low frequencies is enhanced at high flux levels and dominates the other lag components.\par
 \subsection{The distance of the warm absorber}
Previous studies of the complex spectrum of NGC 4051 have found the warm absorber to be located quite close to the black hole and ruled out a torus origin for the gas. Using a simple time-dependent photoionization model on ROSAT data, \cite{nicastro1999} estimate the gas electron density to be $n\sim10^{8}\ \text{cm}^{-3}$, which yields a distance from the X-ray source of $\sim7.4\times10^{15}\ \text{cm}$. \cite{krongold2007} were able to detect a high- and a low-ionization absorbers with XMM-Newton data, and derived a density of $n\sim5.8-21\times10^{6}\ \text{cm}^{-3}$ and $n<8.1\times10^{7}\ \text{cm}^{-3}$ respectively, through time-resolved spectroscopy. These densities lead to an estimate for the distances of $1.3-2.6\times10^{16}\ \text{cm}$ for the high-ionization phase and of $<9\times10^{15}\ \text{cm}$ for the low-ionization phase. More recently with new XMM-Newton observations, \cite{pounds2013} identified a much more complex ionized outflow and propose that the wind is being shocked at a distance of $10^{17}\ \text{cm}$, which places the warm absorber at distances $<10^{17}\ \text{cm}$. Our simulations indicate a maximum effect of the response time of the warm absorber on the X-ray time lags at distances of $\sim0.3-1.0\times10^{16}\text{cm}$, which are comparable to the distance at which the broad-line region is located in this source \citep{denney2009}. Furthermore, these distances correspond to gas densities in the range of $n\sim0.4-3.0\times10^{7}\ \text{cm}^{-3}$, for the two major contributors to the lags, components 1 and 3. Therefore, if we take into account the existing estimates of the location (and density) of the warm absorber and compare them to our results, it appears that the low-frequency X-ray time lags in NGC 4051, measured by \cite{alston2013}, are very possibly affected by the response of the gas to the source variability. \par
%
%
\section{Conclusions}
 \label{section6}
 Working simultaneously with RGS and EPIC-pn data, we performed a detailed analysis using a time-dependent photoionization code in combination with spectral and Fourier spectral-timing techniques. We applied this method to the extensive XMM-Newton archival observations of the bright and highly variable Seyfert 1 galaxy NGC 4051, whose spectrum has revealed a complex multicomponent wind. As a result, we have shown that warm absorbers have the potential to introduce time lags between the most highly absorbed bands relative to the continuum, for a certain range of gas densities and/or distances. The time delay is produced due to the response of the gas to changes in the ionizing source, either by photoionization or radiative recombination.\par
We found that, in the absence of any other lag mechanisms, a soft (negative) lag, of the order of $\sim$ 100 s, is detected when computing the spectral-timing products between the more absorbed energy bands from simulated RGS spectra and a broad continuum band from simulated EPIC-pn spectra, on timescales of hours. Furthermore, we also found that the absorbing gas can likewise produce a time delay between the broader soft and hard bands, both belonging to simulations of EPIC-pn spectra. This happens since the soft band appears to be generally more absorbed, and so even without the higher spectral resolution provided by RGS, we can see the soft band lagging behind the hard, for long timescales. A direct consequence of our results is that understanding the contribution of the recombining gas to the X-ray lags is vital information for interpreting the continuum lags associated with propagation and reflection effects in the inner emitting regions. We have shown that the effects of the warm absorber, in this source, are negligible on short timescales, where the reverberation lag is found. However, on long timescales ($\sim$ hours) the response time of the absorber causes the soft photons to lag behind the hard photons by up to hundreds of seconds. This will have implications for the modelling of the observed hard lags, which have been measured to be only $\sim 50$ s in NGC 4051, since the effect of the warm absorber is to dilute them or even produce dominant soft lags at low frequencies. \par  
The range of gas distances that, under our assumptions, yield a stronger effect of the warm absorber are comparable to the existing estimates for the location of the warm absorber in NGC 4051. In the light of this, we note that it is likely that the warm absorber plays a role in the observed X-ray time-lags in this source. If this is the case, such effects can also help explain the observed soft lags at low frequencies for the low flux segments of the 2009 dataset from NGC 4051. \par
The results we present in this paper are specific to a case study we performed on NGC 4051 and its warm absorber. Assessing the effects of the warm absorber to the lags, through future studies in which we will explore the parameter space, will then allow us to disentangle the contribution from continuum processes and warm absorber response in the lag-frequency and lag-energy spectrum of AGN. As mentioned earlier, one of the problems to study these systems through time-resolved spectroscopy is the timescales involved ($\sim$ minutes to hours), which result in limited photon counts, per time and energy bin. Moreover, there are also other processes playing a role on these timescales, namely continuum processes that produce the hard lag. When doing time resolved spectroscopy the gas response lags are not easily distinguished from the continuum lags, which may cause spurious measurements of the response time. Evaluating the level of uncertainty on these estimates goes beyond the scope of this paper. We stress that using spectral-timing analysis together with a time dependent photoionization model is an extremely powerful method, not only to access the contribution of the warm absorber to the X-ray time lags, but also having the potential to provide important diagnostics on the warm absorber location and gas density, which we will explore in future work. Furthermore, the method can be applied to other sources and warm absorber configurations allowing for a wide range of studies. Indeed, recent work by \cite{kara2016} shows that some AGN with soft lags at low Fourier frequencies are also highly absorbed, highlighting a possible connection between variable absorption and low-frequency lags in other sources. These spectral-timing methods will allow the study of the warm absorber response on even shorter timescales and at higher spectral resolution than were ever possible. With the current dataset on NGC 4051, it is not yet possible to determine whether the lag is associated with the warm absorber, however higher signal to noise grating or calorimeter data may enable this test. Looking further ahead, ATHENA \citep{nandra2013} will allow these methods to be routinely used to study the detailed time response of individual absorption components, allowing us to map AGN outflows in exquisite detail.
%
%
%
%
\begin{acknowledgements}
We would like to thank Jelle Kaastra for providing us with his code as a basis for the development of this method. We would also like to thank Martin Heemskerk for all his help and advice on numerically solving stiff differential equations. Furthermore we would like to thank Gary Ferland, Gerard Kriss, and the participants of the Lorentz Center workshop 'The X-ray Spectral-Timing Revolution' (February 2016), for useful discussions. Simon Vaughan was the PI of the 2009 observation of NGC 4051. 
C. V. Silva acknowledges support from NOVA (Nederlandse Onderzoekschool voor Astronomie). The Space Research Organization of the Netherlands is supported financially by NWO, the Netherlands Organization for Scientific Research. In this work we made use of observations obtained with XMM-Newton, an ESA science mission with instruments and contributions directly funded by ESA Members States and the USA (NASA).  
\end{acknowledgements}
%
\bibliographystyle{bib}
\bibliography{cvsilva}
\end{document}